\documentclass[sigplan,screen]{acmart} 

\AtBeginDocument{%
  }

\setcopyright{rightsretained}
\acmDOI{10.1145/3694848.3694857}
\acmYear{2024}
\copyrightyear{2024}
\acmISBN{979-8-4007-1257-9/24/10}
\acmConference[JENSFEST '24]{Proceedings of the Workshop Dedicated to Jens Palsberg on the Occasion of His 60th Birthday}{October 22, 2024}{Pasadena, CA, USA}
\acmBooktitle{Proceedings of the Workshop Dedicated to Jens Palsberg on the Occasion of His 60th Birthday (JENSFEST '24), October 22, 2024, Pasadena, CA, USA}
\acmSubmissionID{splashws24jensfestmain-p7-p}
\received{2024-06-10}
\received[accepted]{2024-07-12}

\usepackage{amsthm}
\usepackage{amsmath}

\usepackage{amssymb}
\usepackage{xspace}
\usepackage{sansmath}
\usepackage{thmtools}
\usepackage{blkarray}
\usepackage{multicol}
\usepackage{listings}
\usepackage[nameinlink,capitalize,noabbrev]{cleveref}
\usepackage{mathpartir}
\usepackage{wrapfig}
\usepackage[altpo,epsilon]{backnaur}

\newcommand\virgil[1]{\lstinline|#1|\xspace}
\newcommand{\flatten}{\mathsf{flatten}}
\newcommand\myabs[1]{\left| #1 \right|}

\lstdefinelanguage{virgil}{
    morekeywords={class, new, extends, packing, type, def, case, var, return, bool, unboxed, \#packing, \#unboxed, match},
    sensitive=true, 
    morecomment=[l]{//}, 
    morecomment=[s]{/*}{*/}, 
    morestring=[b]" 
}
\definecolor{codegreen}{rgb}{0,0.6,0}
\definecolor{codegray}{rgb}{0.3,0.3,0.3}
\definecolor{codepurple}{rgb}{0.58,0,0.82}
\definecolor{backcolour}{rgb}{0.95,0.95,0.92}
\lstdefinestyle{mystyle}{   
    commentstyle=\color{codegreen},
    keywordstyle=\bfseries\color{magenta},
    numberstyle=\tiny\color{codegray},
    stringstyle=\color{codepurple},
    basicstyle=\ttfamily\color{codepurple}\footnotesize,
    breakatwhitespace=false,         
    breaklines=true,                 
    captionpos=b,                    
    keepspaces=true,                               
    showspaces=false,                
    showstringspaces=false,
    showtabs=false,                  
    tabsize=4,
    rulecolor=\color{black},
    frame=none,
    framesep=1pt
}
\begin{document}

\title{Unboxing Virgil ADTs for Fun and Profit}

\author{Bradley Wei Jie Teo}
\authornote{This work is primarily the honors thesis of Bradley Wei Jie Teo.}
\email{bradleyt@andrew.cmu.edu}
\affiliation{
  \institution{Jane Street}
  \city{New York}
  \state{NY}
  \country{USA}
}

\author{Ben L. Titzer}
\email{btitzer@andrew.cmu.edu}
\affiliation{
  \institution{Carnegie Mellon University}
  \city{Pittsburgh}
  \state{Pennsylvania}
  \country{USA}
}

\renewcommand{\shortauthors}{Bradley Wei Jie Teo and Ben L. Titzer}

\begin{abstract}
  Algebraic Data Types (ADTs) are an increasingly common feature in modern programming languages.
  In many implementations, values of non-nullary, multi-case ADTs are allocated on the heap, which may reduce performance and increase memory usage.
  This work explores annotation-guided optimizations to ADT representation in Virgil, a systems-level programming language that compiles to x86, x86-64, Wasm and the Java Virtual Machine.
  We extend Virgil with annotations: \virgil{\#unboxed} to eliminate the overhead of heap allocation via automatic compiler transformation to a scalar representation, and \virgil{\#packed}, to enable programmer-expressed bit-layouts.
  These annotations allow programmers to both save memory and manipulate data in formats dictated by hardware.
  We dedicate this work as an homage and echo of work done in collaboration with Jens in the work entitled ``A Declarative Approach to Generating Machine Code Tools'', an unpublished manuscript from 2005.
  In fact, this work inherits some syntactic conventions from that prior work.
  The performance impact of these representation changes was evaluated on a variety of workloads in terms of execution time and memory usage, but we don't include it because Jens like semantics and type systems better!
\end{abstract}


\maketitle

\section{Introduction}

\subsection{Algebraic Data Types}
Algebraic Data Types (ADTs), also known as \emph{variants} or \emph{sum types}, are composite types that allow values of that type to take on one of several cases.
Typically, ADTs are used together with pattern matching, a succinct method of discriminating values that ensures matches are exhaustive and type safe.
Once a feature found mainly in functional languages such as Standard ML, OCaml and Haskell, ADTs and pattern matching are an increasingly common feature in modern multi-paradigm languages, such as Rust, Scala and Swift.
In this work, we'll refer to types of this nature as ADTs, runtime instances of them as ADT values (or simply values, if context is clear) and declarations of the cases of an ADT as variants.

\subsection{The Virgil Language}
Virgil\footnote{Virgil is fully open source; the source code, including changes implemented for ADT unboxing can be found at \url{https://github.com/titzer/virgil/}.} is a fully self-hosted systems-level programming language that compiles to multiple targets: WebAssembly (Wasm), x86, x86-64 (Linux and Darwin), and the Java Virtual Machine (JVM).
Virgil provides higher-level features such as classes, first-class functions and closures \cite{v3}.
A key feature of Virgil is that programs can execute arbitrary code at compilation time via a built-in interpreter for the entire language.
The initialized state of the program, including a graph of live objects, is included as the initial heap image in the emitted executable.
The compiler employs a number of optimizations on the combined code and data of a program, including reachable members analysis, constant-propagation, inlining, and a suite of classical compiler optimizations.

The first published works on Virgil targeted embedded systems.
In fact, the first paper entitled ``Virgil: Objects on the Head of a Pin'' appeared in OOPSLA 2006~\cite{v3object} and targeted microcontrollers, a key interest of Jens at the time, and UCLA was consumed with the sensor network Zeitgeist.

Today, Virgil also features automatic memory management, either reusing the target VM's object model and garbage collector (GC), such as when targeting the JVM, or using its own precise tracing, Cheney-style copying garbage collector for native and Wasm targets.
Importantly, until today, Virgil's static type system and compilation strategy has ensured a static separation between locations that can store references to heap objects and those that cannot.
This allows the runtime system to distinguish references by consulting a stackmap during stackwalking and bitmaps when tracing objects.
In this paper, we introduce unboxing techniques that transparently unify reference and non-reference values when/if the target and runtime system supports tagged pointers, which allows a single location to store either references or primitive values that are distinguished with tag bits.

\subsection{ADTs in Virgil}
ADTs are a relatively recent addition to Virgil and have not been described yet in its published literature.
As in most functional languages, ADTs in Virgil are \emph{value types}; values are immutable and equality comparison is structural.
One wrinkle is that Virgil requires every type have a default value\footnote{A variable takes on its type's default value when it is not explicitly initialized.}.
Virgil defines the default value of an ADT to be an instance of the first declared variant with all fields set to their respective default values.
Another difference is that not only is each ADT definition associated with \emph{one} type; but each variant defines a name for a more-precise ``child'' type, which is a subtype of the enclosing ADT.
Virgil further blends object-oriented and functional programming paradigms by allowing ADTs to define methods that are dispatched based on the value's case, as if the ADT defined a superclass and the cases were subclasses\footnote{In fact, to reuse much compiler logic, ADTs are desugared to classes early in compilation.}.
This allows some syntactic conveniences and reduces friction when refactoring a type from being a class hierarchy to being an ADT or vice versa.

\begin{figure}
\begin{lstlisting}[language=virgil]
type Option<T> {
  case None;
  case Some(val: T);

  def val() -> T { return Some<T>.!(this).val; }
  def isNone() -> bool { return None.?(this); }
  def isSome() -> bool { return tag == 1; }
}
\end{lstlisting}
\label{fig:adt-example}
\caption{An generic option type written in Virgil, showcasing features of Virgil ADTs, including type parameters, cases with fields, and methods.}
\end{figure}

An illustration of the syntax of ADTs in Virgil is shown in \cref{fig:adt-example}, which defines a generic \virgil{Option<T>} type.
We see that type definitions may be parameterized over type parameters, i.e. be \emph{polymorphic}.
We also see that values can be casted (e.g. \virgil{Some<T>.!()}, which may produce a runtime \virgil{TypeCheckException}) and queried (e.g., \virgil{None.?()}, which produces a boolean value).
Further we see that values have fields \emph{tag}, an integer, and a \emph{name}, a string, which can be accessed using \virgil{.tag} and \virgil{.name}.
These fields can come in handy, e.g. to index into an array based on the tag or to print the name of the variant, without writing a match.

\subsection{Motivating Unboxing}

ADTs bring new expressiveness to Virgil, allowing convenient data type definitions, matching, and polymorphism.
Referential transparency and structural equality allow the compiler implementation freedom and insulate programs from machine details like pointer indirections and heap allocations.
Yet as a new basic abstraction mechanism, the representation of ADT values can make a large difference in program memory usage and performance.
Moreover, for a systems language where resource usage should be predictable and low, programmers should have insight and control into how ADT values are represented, especially since different representations have different space/time tradeoffs.
Hence, we want the compiler to choose representations for ADTs as scalars efficiently, using a generalizable but target-sensitive algorithm.

Prior to this paper, ADT values in Virgil were represented in one of three ways, depending on their number of variants and fields:
\begin{enumerate}
\item Instances of classes allocated on the heap, desugared by the parser into classes at the AST level. This was the default for all but trivial ADTs, and uses the same record infrastructure to represent ordinary classes.
	\item A single unsigned integer, if all variants are trivial. This degenerates ADTs into enums when possible.
	\item Individual scalars\footnote{We use \emph{scalar} to refer to values that will not be split further at lower levels of the compiler, with the exception of numeric lowering on 32-bit targets. A rough approximation of a scalar would be a register value on x86 or a variable in the JVM.}, for \emph{explicitly unboxed}\footnote{There is a compiler flag to unbox all non-recursive single-variant ADTs that normalize to fewer than a given number of fields. Otherwise, unboxing is done for single-variant ADTs with the \lstinline|#unboxed| annotation.} single-variant ADTs. Such ADTs are represented as multiple separate values without a tag.
\end{enumerate}

The first strategy is easy to implement in a compiler that already has support for classes, as desugaring of variants into classes reduces the amount of variant-specific logic during compilation (\cref{fig:desugaring}).
However, this limits the efficiency of ADT representations, as it means that all non-empty, multiple-variant ADTs must incur a heap allocation on initialization and an indirection on field access.
This can hurt performance, not only due to the overhead of allocation, but also due to the increase in garbage collection pressure due to more and potentially larger allocated objects.

In this work (because Jens has taught us well that ADT values should be referentially transparent) we explore \emph{unboxing} ADT representations where the compiler represents ADT values as a combination of scalars and tags.
We give a syntax for unboxing annotations and describe how the compiler can transparently--i.e. without changing program semantics--compile programs to avoid heap allocations.

In addition to unboxing, we explore annotations for \emph{packing} several fields of a variant into a single scalar, allowing smaller representations, controlling representation down to the bit level.
The syntax we propose follows prior work by Titzer and Palsberg in ``A Declarative Approach to Machine Code Tools''~\cite{DeclMach}.

\begin{figure}
\begin{lstlisting}[language=virgil]
  type T {
    case A(x: int);
    case B(y: float);
  }
  // desugars to
  class T { }
  class T.A(x: int) extends T { }
  class T.B(y: float) extends T { }
\end{lstlisting}
\caption{An example of a Virgil ADT definition and its internal desugaring to classes.}
\label{fig:desugaring}
\end{figure}

\section{Stuff Jens Knows}

We posit that Jens has an encyclopedic knowledge of programming languages and bring to attention a selection of related work.
Jens knows that work has been done on unboxing ADTs through automated monomorphization and specialization in Haskell~\cite{hall96}.
Further, Jens knows that MLTon implements ADT unboxing for Standard ML and can also specialize, as it is a whole-program compiler~\cite{mlton}.
However, it does not provide a mechanism for programmer annotations.
Jens has seen recent work in this area that does add annotations~\cite{UnboxedData}.

Jens knows that Rust, which compiles to LLVM IR, represents ADTs as tagged values.
For unboxed ADTs, they become an LLVM structure type with a tag and an array of associated bytes (the size of the largest \lstinline|enum| case).
This is feasible because Rust is a non-garbage-collected language which does not need to differentiate between reference and non-reference types at runtime.
Virgil's JVM target forbids the representation of ADTs as raw bytes of data in the early stages of the compiler\footnote{Representing ADTs as raw, packed bytes without consideration for scalars would mean fields may be inadvertently split across variables, making their access costly.}.

Jens knows that there is an open RFC on unboxing ADTs in OCaml \cite{ocamlrfc22} with annotations.
Recently, a version of the OCaml compiler supports unboxing where unboxed cases have the same representation as their fields \cite{chataing2023unboxed}.
There are certain limitations that limit the power of this optimization.
OCaml requires a uniform representation of values because it must compile generic code without concrete types.
Virgil has a whole-program compiler that performs a monomorphization pass, so concrete types of all values are known during normalization and would be delighted to learn that this work exploits this.

Jens knows that Rust performs bit-level tag packing using \emph{niche optimization}.
Previously, Rust would only perform niche optimizations on ADTs with several nullary cases and a single non-null variant containing a `niche' \cite{niche}.
Recently, improvements in Rust allow for non-nullary cases that start or end with data, as long as they do not interfere with the niche \cite{niche2023}.
Swift supports a similar optimization, also enabling several non-nullary cases \cite{swift}.

Jens knows that Zig supports packed structs and packed unions, enabling the expression of bit-level layouts but without flexibility on tag location \cite{zigdoc}.
A proposed feature (that has no current plans for implementation) would enable user-specified tagging for packed unions \cite{zigrfc}.

Memory layout, a related problem to unboxing, has important relation to this work.
Jens knows that Ribbit is a domain-specific language for memory layouts that targets LLVM IR \cite{bitstealing}.
It is flexible enough to express the default compiler representations for OCaml and Rust ADTs.
Recently, Dargent appeared, a description and refinement language for memory layouts in Cogent~\cite{cogent}.
It allows for programmer-specified memory layouts of boxed values allocated on the heap.
Virgil supports byte-level layouts that act as views over byte arrays \cite{v3layouts}.
Jens knows that C\# supports a similar feature for control over a struct's memory layout, using the \lstinline|FieldOffset| annotation \cite{csharplayout}.
A recent update to Odin adds support for bit-level layouts, which it calls \emph{bit fields}, backed by a user-specified type \cite{odin}. Odin's bit fields also supports specifying the endianness of fields.

\section{The Virgil Compiler}


\begin{figure}
	\includegraphics[width=2in]{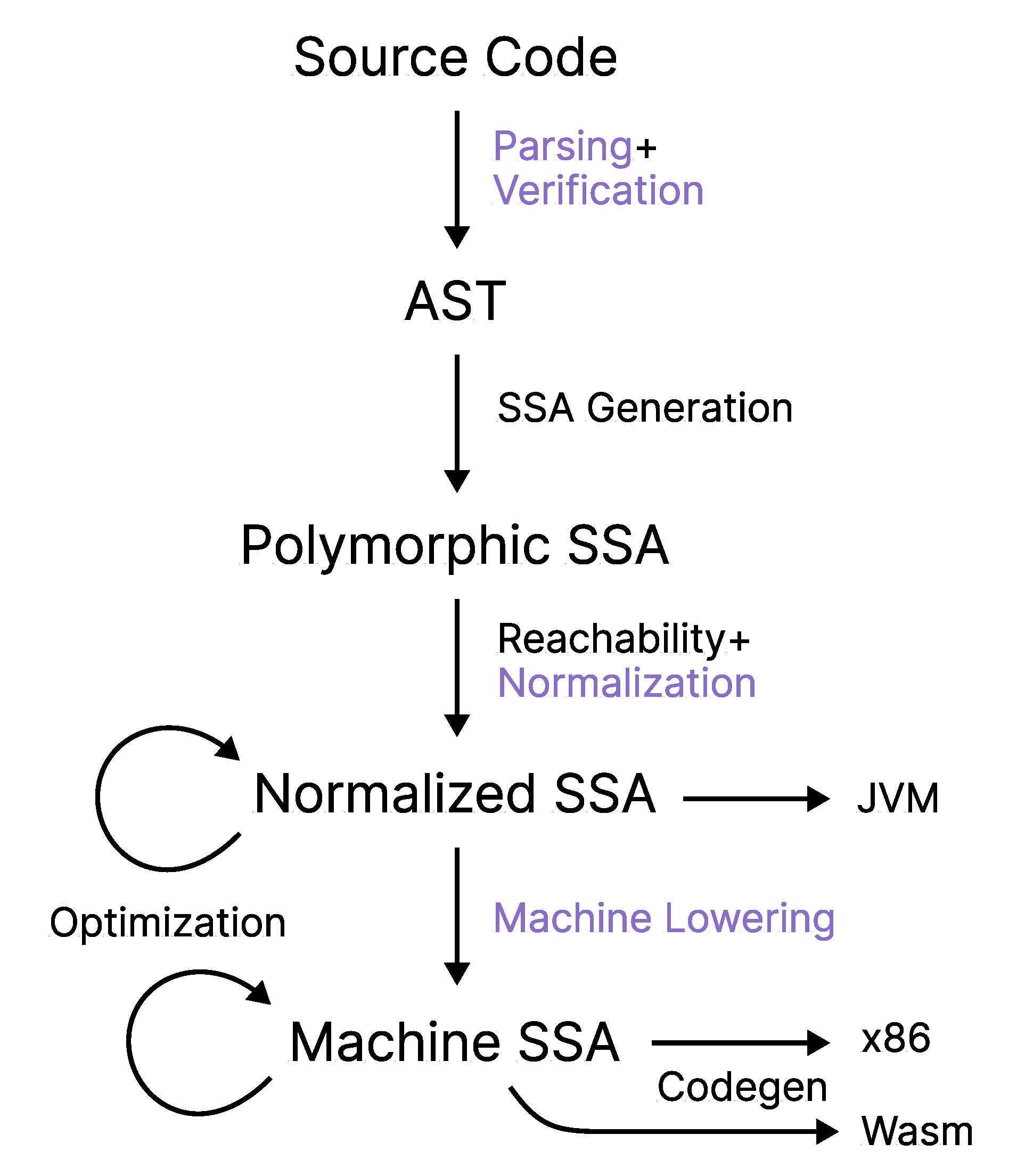}
	\caption{A diagram of the Virgil compiler's phases. The largest changes from this work are in purple.}
	\label{fig:compiler}
\end{figure}
        
Jens has taught us to conceptualize compilation as a series of translations.
The Virgil compiler undergoes several phases to lower programs into machine code (\cref{fig:compiler}). \begin{enumerate}
	\item \emph{Parsing and Verification}: Source to AST;
	\item \emph{SSA Generation}: AST to Polymorphic SSA;
	\item \emph{Reachability and Normalization}: Polymorphic SSA to Normalized SSA;
	\item \emph{Machine Lowering}: Normalized SSA to Machine SSA;
	\item \emph{Code Generation}: Machine SSA to Machine Code.
\end{enumerate}

The phases differ somewhat between different targets; for instance, the JVM target skips machine lowering and instead directly translates Virgil object operations into JVM object operations.
To simplify code generation for Virgil's fixed bit-width integer operations, \emph{numeric lowering} occurs during the machine lowering phase and decomposes integer values and operations wider than the machine width to multiple machine words and operations.

\subsection{Reachability and Normalization}
The Virgil compiler performs specialization and dead code elimination as part of its compilation strategy via a reachability analysis.
During reachability, the live variables, fields and instructions are determined by a recursive search over the program.
Here, the compiler determines if fields always take on constant values or are never used and uses information during normalization to optimize code, remove dead fields, and otherwise choose efficient representations~\cite{v3object}.

The Static Single-Assignment (SSA) IR of Virgil contains type information.
Normalization of this SSA is directed by type information in SSA and the information from reachability.
At this phase, complex types such as tuples and closures are flattened to multiple scalars.
Any representation that consists of more than one scalar is referred to as \emph{multi-scalar}, though the IR is rewritten to refer to the constituent scalars whereever possible\footnote{The only exception being returns in the SSA IR, which temporarily pack multiple scalars into a tuple that are then unpacked at call sites.}.

\subsection{Compiler Refactorings to Support Unboxing}
The introduction of unboxed ADTs involves changes in several phases of compilation.
\begin{itemize}
	\item \textbf{Parsing.} The parser was modified to accommodate the bit-level layout syntax and the new annotations.
	\item \textbf{Verification.} The verifier now verifies that the packing declarations are semantically possible. This is described in \cref{sec:verification}.
	\item \textbf{Normalization.} In this phase, polymorphic code is translated to monomorphic code with normalize representations; tuples and closures are flattened, fields are replaced with sub-fields, and virtual dispatches are optimized.
          We made substantial changes to this phase to support new representations.
	 In this phase we introduce a new \emph{solving} algorithm which searches for good representations for unboxing ADTs.
	\item \textbf{Machine Lowering.} This phase has to be aware of the new packed representation of types in order to generate precise stackmaps.
	\item \textbf{Runtime.} With the addition of tagged pointers, the garbage collector must be changed together with the compiler. The target specifies the bit patterns for reference and non-reference tagged pointers, and the GC must be modified to handle tagged pointers, distinguishing between reference and non-reference values, and masking out the correct bits to extract the underlying reference.
\end{itemize}

\section{A Language for Bit-Level Layouts}
It is intractable to automatically derive an optimal data representation, for many reasons.
In particular, program behavior may favor one representation under one workload and another under a different workload.
In this paper we introduce a syntax called \emph{packings} that allows programmers control over bit-level representation, so that programmers can tailor data representations to their needs.
We embed this simple language within Virgil to allow specifying bit-level packings for ADTs.

\subsection{Syntax}
Our syntax gives programmers a range of options for how explicitly they write a bit-level layout.
In the most explicit option, one writes an \emph{extended binary literal}, which is a pattern that specifies the meaning of every single bit, including bits from ADT fields, specifying fields bit-by-bit from the most-significant to least-significant using either \virgil{0}, \virgil{1}, \virgil{?}, or the first character of their name\footnote{If this is ambiguous, such as when two fields share the first character, an error is reported during verification.}.
Bits with \virgil{?} indicate the compiler may choose the bit value (as part of solving, discussed later) and cannot be observed by the program.
Less explicitly, packing expressions can also include the \emph{concatenation} of other packing expressions using \lstinline|#concat| and the application of packing declarations.
\begin{figure*}
\begin{lstlisting}[language=virgil]
packing Float16(sign: 1, exp: 5, frac: 10): 16 = 0b_seeeeeff_ffffffff;
packing Float32(sign: 1, exp: 8, frac: 23): 32 = 0b_seeeeeee_efffffff_ffffffff_ffffffff;

packing TwoFloat16s(s1: 1, e1: 5, f1: 10, s2: 1, e2: 5, f2: 10): 32
	= #concat(Float16(s1, e1, f1), Float16(s2, e2, f2));
\end{lstlisting}
\caption{A representation of IEEE 754 floating-point numbers using our packing declaration syntax, followed by an example of packing application and concatenation.}
\label{fig:float-eg}
\end{figure*}

\begin{figure*}
	\begin{bnf*}
	\bnfprod{packing-expr}{\bnfes \bnfor \bnfpn{bit-layout} \bnfor \bnfpn{const}  \bnfor \bnfpn{ident} \bnfsp \bnfts{(} \bnfsp \bnfpn{packing-expr-list} \bnfsp \bnfts{)} } \\
	\bnfmore{\bnfts{\#concat} \bnfsp \bnfts{(} \bnfsp \bnfpn{packing-expr-list} \bnfsp \bnfts{)} \bnfor \bnfts{\#solve} \bnfts{(} \bnfsp \bnfpn{packing-expr-list} \bnfsp \bnfts{)}} \\
	\bnfprod{packing-expr-list}{\bnfes \bnfor \bnfpn{packing-expr} \bnfor \bnfpn{packing-expr} \bnfsp \bnfts{,} \bnfsp \bnfpn{packing-expr-list}} \\
	\bnfprod{packing-decl}{\bnfts{packing} \bnfsp \bnfpn{ident} \bnfsp \bnfts{(} \bnfsp \bnfpn{param-list} \bnfsp \bnfts{)} \bnfsp \bnfts{:} \bnfsp \bnfpn{int} \bnfsp \bnfts{=} \bnfsp \bnfpn{packing-expr} \bnfsp \bnfts{;} } \\
	\bnfprod{param-list}{\bnfes \bnfor \bnfpn{ident} \bnfsp \bnfts{:} \bnfsp \bnfpn{int} \bnfor \bnfpn{ident} \bnfsp \bnfts{:} \bnfsp \bnfpn{int} \bnfsp \bnfsp \bnfts{,} \bnfsp \bnfpn{param-list}} \\
	\bnfprod{packing-annot}{\bnfts{\#packing} \bnfsp \bnfts{(} \bnfsp \bnfpn{packing-expr-list} \bnfsp \bnfts{)} \bnfor \bnfts{\#packing} \bnfsp \bnfpn{packing-expr}} \\
	\bnfprod{bit-layout}{\bnfts{0b} \bnfsp \bnfpn{packing-bits} } \\
	\bnfprod{packing-bits}{\bnfes \bnfor \bnfpn{packing-bit} \bnfsp \bnfpn{packing-bits} } \\
	\bnfprod{packing-bit}{\bnfts{0} \bnfor \bnfts{1} \bnfor \bnfpn{char} \bnfor \bnfts{?}}
	\end{bnf*}
	\caption{Syntax for packing expressions and declarations, in Backus-Naur form.}
	\label{fig:packing-syntax}
\end{figure*}

Packings can appear as top-level declarations, allowing them to be reused by name throughout the program, and also as annotations on ADTs.
When packings are attached to variants, \lstinline|#solve| expressions instruct the compiler to figure out a way to pack fields for under-specified layouts.
\lstinline|#solve| expressions cannot appear in packing declarations, which must be fully specified.
A full description of the syntax of packing expressions can be found in \cref{fig:packing-syntax}.

\subsection{Static Verification} \label{sec:verification}
During semantic analysis, the compiler verifies that packing declarations are well-formed by checking the sizes of packing expressions.
Define the judgment $\Delta, \Gamma \vdash e:n$ to be that packing expression $e$ has size at most $n\in \mathbb{N}$ in the contexts $\Gamma$ and $\Delta$.
$\Delta$ is the context containing the packing declarations defined in the program. Expressions that are too short will be padded with zeros.
We restrict $n$ to be the size of the largest possible scalar ($64$ bits on 64-bit targets).
\begin{figure*}
\begin{mathpar}
	\inferrule{\Delta, \Gamma\vdash e:n \\ n\leq n'}{\Delta, \Gamma\vdash e:n'}

	\inferrule{\myabs{b}=n}{\Delta, \Gamma\vdash b:n}

	\inferrule{\Gamma(f)=n}{\Delta, \Gamma\vdash f:n}

	\inferrule{\Delta, \Gamma\vdash e_i:n_i}{\Delta, \Gamma\vdash\mathsf{concat}(\overline{e_i}):\sum n_i}

	\inferrule{\Delta, \{\overline{x_i\mapsto n_i}\}\vdash e:n}{\mathsf{packing}\; p(\overline{x_i:n_i}):n=e \; \mathsf{ok}}

	\inferrule{\mathsf{packing}\; p(\overline{x_i:n_i}):n=e\in\Delta \\ \Delta, \Gamma\vdash e_i:n_i}{\Delta, \Gamma\vdash p(\overline{e_i}):n}
\end{mathpar}
\caption{A subset of the static rules for packing expressions.}
\end{figure*}

ADTs can be polymorphic in Virgil, and thus variants can have fields of polymorphic type, and thus unknown bitwidth.
Thus it is not possible to verify all packing annotations at typechecking time.
For this reason, packing annotations on variants are not verified until normalization, since concrete types of fields are known after monomorphization.

\section{Solving the Unboxing Problem}

\subsection{Conditions for Unboxing}
We provide the \lstinline|#unboxed| annotation for the programmer to annotate a variant as unboxed.
Additionally, single-variant ADTs that normalize to a small number of scalars are automatically unboxed.

However, there are two situations in which variants must be boxed:
\begin{itemize}
	\item If a variant is recursive (or mutually recursive), our algorithm will not unbox it.\footnote{This is more restrictive than necessary; it should be possible to unbox recursive variants if all the recursive \emph{mentions} of that type are boxed.}
	\item If a variant value is captured in a closure (for example, a method \lstinline|T.f| is referred to without being called), our algorithm will box \lstinline|T|, as Virgil represents closures as a code pointer and environment pointer pair\footnote{Another option would be to transparently box the value at closure creation time, but we decided against introducing implicit heap allocations here.}.
\end{itemize}

\subsection{Scalar and Interval Assignment}
We can think of the unboxing problem as two different problems: an assignment of normalized fields to scalars (\emph{scalar assignment}), and an assignment of fields to \emph{bits} in those scalars (\emph{interval assignment}).
However, these two problems are deeply intertwined -- it is not possible to determine if a scalar assignment is valid without trying to find an interval assignment on those scalar assignments -- and so they must be solved together.
Another benefit is that performing an explicit scalar and interval assignment, as opposed to just performing interval assignment on one long bit layout, prevents splitting fields across multiple scalars.

Unlike prior work, Virgil's monomorphization and normalization strategy allows scalar assignment to use an unbounded number of scalars to represent an ADT value.
This allows a naive solution where unboxing and packing use as many scalars as necessary to store the fields for \emph{all} variants, plus a tag.
Obviously, this is very inefficient, so a heuristic for unioning fields across variants is necessary.
Our algorithm for solving the unboxing problem uses recursive backtracking, improved with various heuristics, to minimize the number of scalars and cost of field access.

\begin{figure*}
	\centering
	\includegraphics[width=0.9\textwidth]{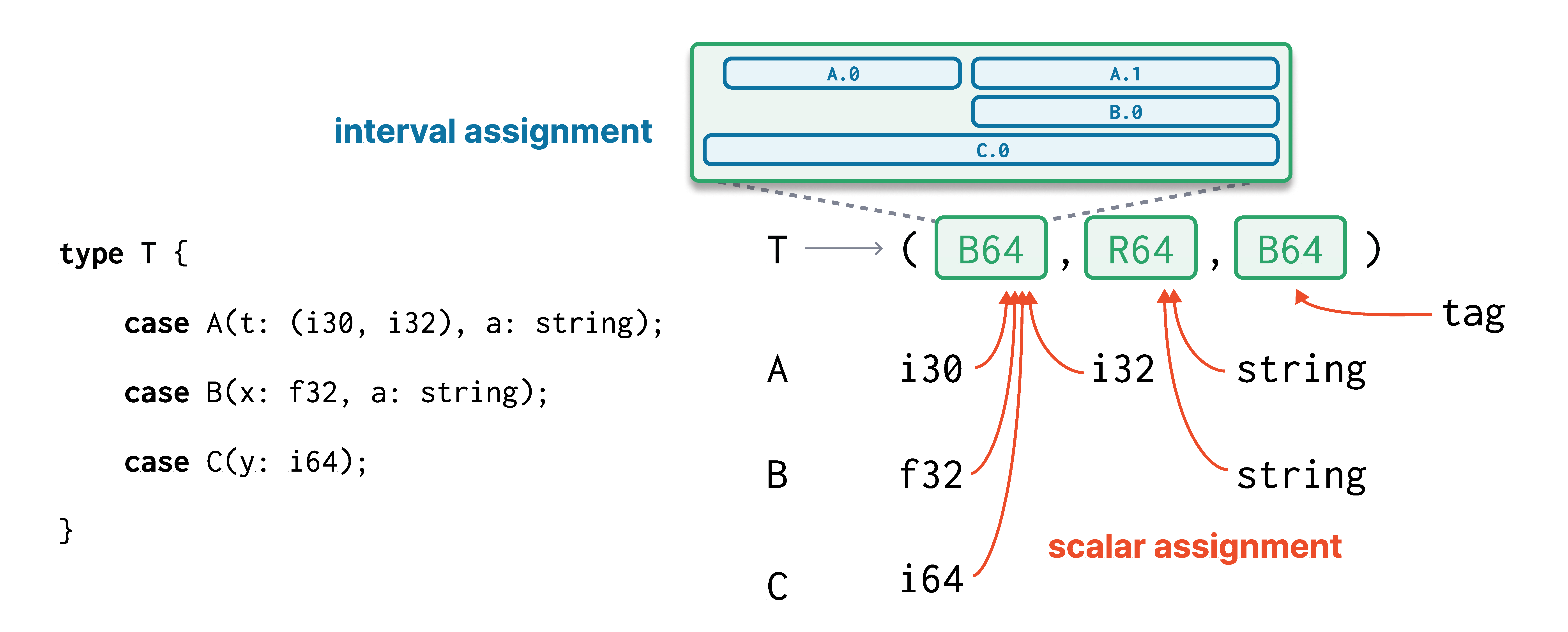}
	\caption{An illustration of scalar and interval assignment.}
	\label{fig:assignment}
\end{figure*}

\subsection{Scalar Kinds}

To allow the unboxing and packing algorithms to reason about machine representations, we introduce the idea of a \emph{scalar kind}.
A scalar kind is similar to the notion of a \emph{register class} in a compiler backend, but we avoid this term to prevent confusion.
The valid scalar kinds are \texttt{B32}, \texttt{B64}, \texttt{R32}, \texttt{R64}, \texttt{Ref}, \texttt{F32} and \texttt{F64}, which constitute the union of scalar kinds over all supported targets.\footnote{In the future, we may include \texttt{B128} as a scalar kind to support XMM registers on x86-64 with SIMD, or Wasm SIMD.}

During normalization, the target specifies a mapping \texttt{GetScalar} between normalized types and \emph{sets} of scalar kinds.
This set represents the kinds of scalars that can physically store a value of the type.
For instance, on the x86-64 target, a value of type \lstinline{u2} could inhabit a \texttt{B64}, \texttt{F64}, or \texttt{R64} register.\footnote{If we enable packed references and tagged pointers with alignment, the compiler may use low-order bits to store fields or the tag}.
Yet for the JVM target, an \lstinline{Array<byte>} instance can only occupy a \texttt{Ref} scalar, since references in Java are opaque.

Scalar kinds allow the target express the unifications that are possible between different fields.
They also allow the algorithm to reason about which scalar kinds are preferential: for instance, mapping \lstinline{int}'s to $\{\texttt{B32}\}$ and \lstinline{float}'s to $\{\texttt{B32}, \texttt{F32}\}$ will have their merged value live in a integer register, since the intersection is $\{\texttt{B32}\}$.
The distinction between the reference and non-reference scalar kinds also guide the compiler in later phases, when it builds the stackmap for garbage collection.

\subsection{Recursive Backtracking}
In order to create an assignment of fields to scalars, we perform recursive backtracking on the assignment of each scalar.
As we iterate over the fields present in the variant, we build up a representation by assigning them to available scalar slots.
When all fields from all cases have been assigned, we check if the assignments have a valid packing that is also \emph{distinguishable} (discussed next).
In order to limit the time taken to find a good solution, the number of solving steps is bounded\footnote{We set the bound high enough that the solver will at least find the trivial solution where the ADT is the union of all cases' fields.}.

\subsection{Distinguishability: Explicit and Implicit Tagging}
Values of different variants must be distinguishable at runtime, both for pattern matching and GC tracing.
It is insufficient merely to pack the fields for each variant; for multi-variant ADTs, we need enough information in the packed representation to distinguish between cases.
After interval assignment, the representation tracks the unassigned bits in the scalar that we can use for this purpose.

\begin{figure}
	\includegraphics[width=3.1in]{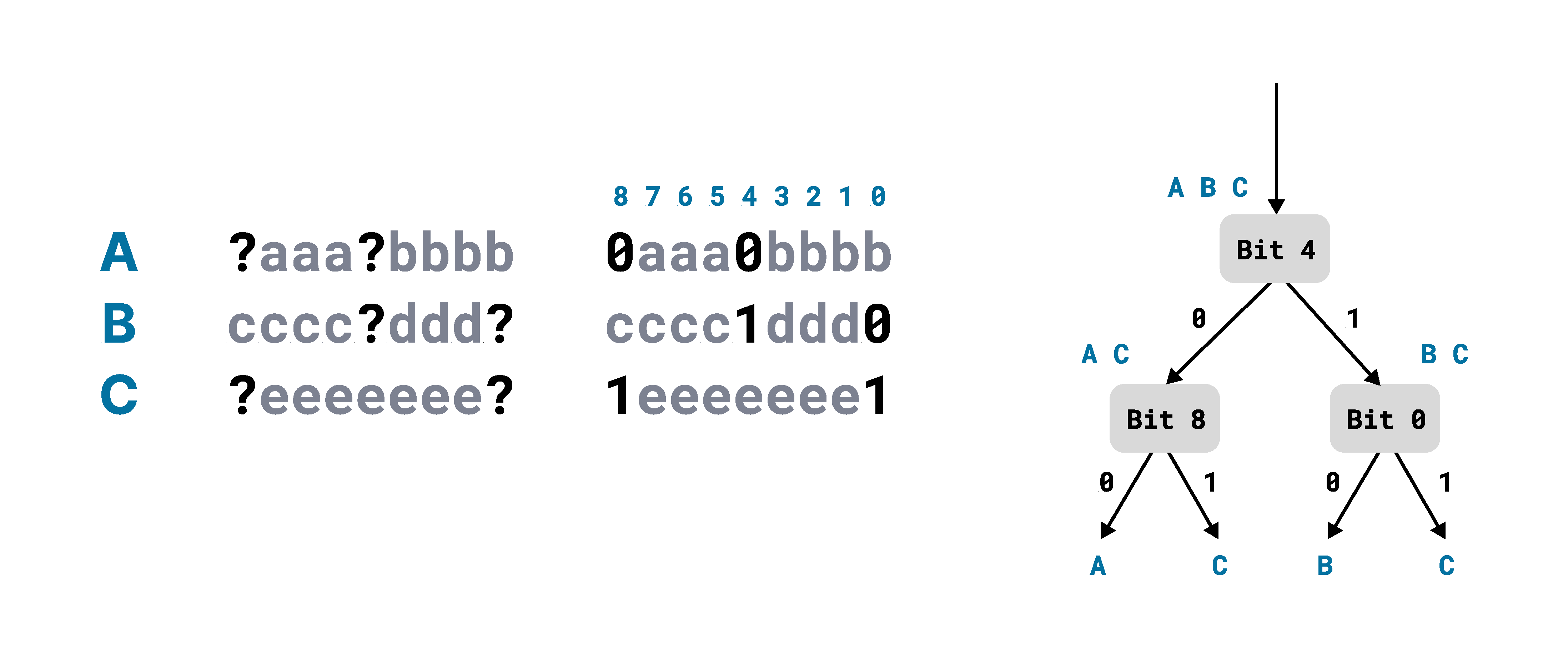}
	\caption{Decision tree derivation.}
	\label{fig:tree}
\end{figure}

We identified two options for distinguishing cases.
\begin{itemize}
	\item \textbf{Explicit Tagging.} If there are sufficiently many aligned, unassigned and contiguous bits after interval assignment, we can encode the case's tag as a field in that interval. If this isn't possible, we can append a new scalar that acts as the variant's tag.
	\item \textbf{Deriving a Decision Tree.}
          \footnote{Not yet implemented, only verifying the distinguishability of scalars.} There may be sufficient information in the concrete bits of packing patterns, or enough unassigned bits, to distinguish between cases, but they are not all correctly aligned or chosen.
          If this is true, we can distinguish between the various cases by carefully choosing assignments for unassigned bits and constructing a decision tree; each node in the decision tree represents sets of cases that have yet to be distinguished; each node splits on a specific bit position (\cref{fig:tree}).
          This is directly derived from work with Palsberg in~\cite{DeclMach}.
	
\end{itemize}

\subsection{Heuristics}
The afore-mentioned recursive backtracking algorithm is extremely inefficient on its own; we need the use of heuristics in order to make the packing problem tractable.
Additionally, we need some way to \emph{score} solutions, so that the solver is able to pick between multiple valid solutions. Our score is a function of several parameters: \begin{itemize}
	\item \textbf{Number of scalars.} We penalize solutions that use more parameters, as we would like to encourage smaller representations.
	\item \textbf{Access cost.} When possible, we would also like fields to occupy their own scalar, as access cost is reduced. This factor helps distribute fields across scalars, especially for variants with one large case.\footnote{One additional contributor to access cost is casting cost on the JVM: if references are represented as \lstinline|java.lang.Object|s, then their casting incurs a bytecode instruction. We don't yet consider this as a cost in the current implementation.}
	\item \textbf{Presence of explicit tag.} This is essentially the access cost of a variant's tag.
\end{itemize}

\subsection{Flattening Packing Annotations}

\begin{figure*}
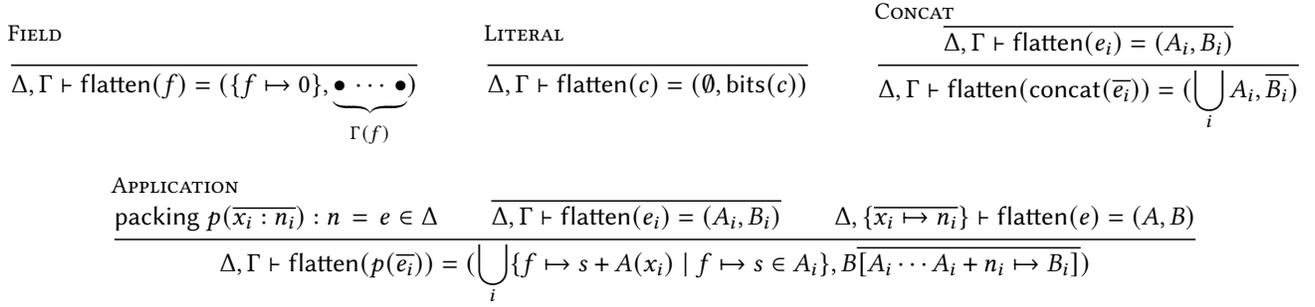

	\begin{mathpar}
		\inferrule[Field]{ }{\Delta, \Gamma\vdash\flatten(f) = (\{f \mapsto 0\}, \underbrace{\bullet\,\cdots\,\bullet}_{\Gamma(f)})}
	
		\inferrule[Literal]{ }{\Delta, \Gamma\vdash\flatten(c) = (\emptyset, \mathsf{bits}(c))}
	
		\inferrule[Concat]{\overline{\Delta, \Gamma\vdash\flatten(e_i) = (A_i, B_i)}}{\Delta, \Gamma\vdash\flatten(\mathsf{concat}(\overline{e_i}))=(\bigcup_i A_i, \overline{B_i})}
	
		\inferrule[Application]{\mathsf{packing}\; p(\overline{x_i: n_i}):n\,=\,e \in \Delta \\ \overline{\Delta, \Gamma\vdash\flatten(e_i) = (A_i, B_i)} \\ \Delta, \{\overline{x_i\mapsto n_i}\} \vdash \flatten(e)=(A, B)}{\Delta, \Gamma\vdash \flatten(p(\overline{e_i}))=(\bigcup_i \{f \mapsto s+A(x_i) \mid f\mapsto s\in A_i\}, B\overline{[A_i \cdots A_i + n_i \mapsto B_i]})}
	\end{mathpar}
	\caption{Rules for flattening packing expressions into patterns and assignments.}
	\label{fig:rules-flatten}
\end{figure*}

Given a \lstinline|#packing| annotation, we must convert it into a series of scalars and interval assignments that is recognized by the solver.
Each \emph{flattened} packing expression is a pair containing a bit pattern\footnote{Each packing bit is either 0, 1, assigned ($\bullet$) or unassigned ($?$).} and a list of interval assignments. This flattening takes place according to the rules in \cref{fig:rules-flatten}.
The context $\Gamma$ stores a mapping of parameters or fields to bit widths; we assume an ambient context $\Delta$ containing all packing declarations in the program.
\begin{itemize}
	\item The rule for flattening bit layouts is omitted, but is as expected: for example, flattening the pattern \lstinline|0b_00aa_bb11| becomes $(\{a\mapsto 4, b\mapsto 2\}, 00\bullet\bullet\bullet\bullet 11)$, since $a$ is at bit position $4$ and $b$ at $2$.
	\item The rule for application states that we flatten the packing declaration's expression with placeholder fields representing parameters.
          We flatten all the arguments, insert their patterns into the appropriate spots, and shift the assignments by their positions in the full expression.
\end{itemize}

\section{Code Normalization}
\begin{figure}
	\centering
	\includegraphics[width=3.5in]{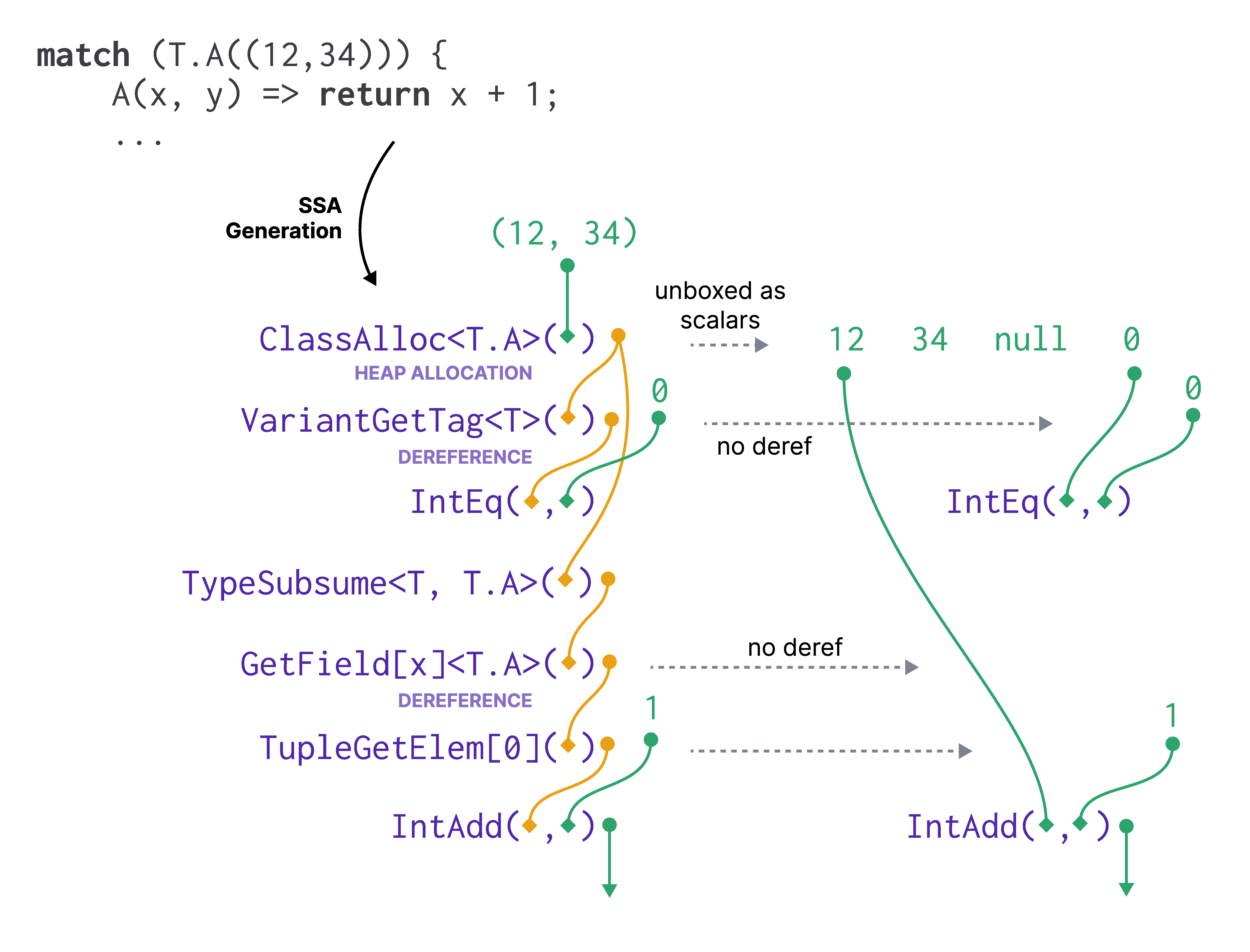}
	\caption{An illustration of SSA rebuilding.}
	\label{fig:normalization}
\end{figure}

After solving for a well-scoring unboxing and packing solution, the SSA\footnote{Although the code of each method is in SSA form, no interprocedural SSA constructs are used.} IR must be rewritten in order to make use of this new representation (\cref{fig:normalization}).
This takes place during the \emph{normalization} phase, which rewrites the entire program to flatten all other types that have multi-scalar representations.

In the front-end of the Virgil compiler, ADTs are desugared to classes, so they share some of the same SSA operations with classes.
There are several SSA operations that can operate or create these ADT values:
\begin{itemize}
	\item \lstinline|ClassAlloc|, which creates a value of the specified ADT case from its constituent fields;
	\item \lstinline|VariantGetField|, which extracts a specific field from an ADT value;
	\item \lstinline|VariantGetTag|, which extracts a numeric tag from an ADT value;
	\item \lstinline|VariantReplaceNull|, which replaces a possibly null reference with the default value of that variant. This is a minor compiler quirk, as the default value of a class is \lstinline|null|, but ADTs are non-nullable.
\end{itemize}

\subsection{ADT Operations}
We present a simplified but illustrative example of the normalization process for ADTs.
Suppose our pre-normalization typed SSA supports the following types:
\[
	\tau \ \ ::=\ \ \mathsf{int} \quad|\quad (\overline{\tau_i}) \quad|\quad \mathsf{class}(\tau)
\]
while the post-normalization typed SSA expects the types
\[
	\rho\ \ ::=\ \ 	\mathsf{int} \quad|\quad \mathsf{class}(\rho) \quad|\quad (\overline{\rho_i}).
\]

Consider an ADT of the form 
$\tau = \mathsf{adt}(\overline{c_i \hookrightarrow \tau_i})$, in which the desugaring process has synthesized the associated types $\tau_{c_i} = \mathsf{class}(\tau_i)$ and $\tau_c=\mathsf{class}(\cdot)$.
Operations on the variant $\tau$ have been generated as typed SSA operations on class types.

As described in the previous section, we create a \emph{variant norm} for $\tau$ that contains information about how the fields and tag are represented ($\rho_\text{field}$ and $\rho_\text{tag}$).
We normalize types as in \cref{fig:type-norm}.
\begin{figure*}
\begin{mathpar}
	\inferrule{ }{\mathsf{int} \rightsquigarrow \mathsf{int}}

	\inferrule{\overline{\tau_i \rightsquigarrow \rho_i}}{(\overline{\tau_i}) \rightsquigarrow (\overline{\rho_i})}

	\inferrule{\mathsf{class}(\tau) \text{ boxed} \\ \tau\rightsquigarrow\rho}{\mathsf{class}(\tau) \rightsquigarrow \mathsf{class}(\rho)}

	\inferrule{\mathsf{class}(\tau) \text{ unboxed variant}}{\mathsf{class}(\tau) \rightsquigarrow (\rho_\text{field}, \rho_\text{tag})}
\end{mathpar}
\caption{Normalization of types.}
\label{fig:type-norm}
\end{figure*}

Given a pre-normalization typed SSA with instructions \begin{align*}
	p \ \ ::=\ \  \cdots \ \ |\quad &x=\text{\ttfamily Const}(v) \\
	|\quad &x=\text{\ttfamily ClassAlloc}\langle\tau\rangle(y) \\
	|\quad &x=\text{\ttfamily GetContents}\langle\tau\rangle(y) \\
	|\quad &x=\text{\ttfamily GetTag}\langle\tau\rangle(y)
\end{align*}
and a post-normalization typed SSA with instructions \begin{align*}
	q \ \ ::=\ \  \cdots \ \ |\quad &x=\text{\ttfamily Const}(w) \\
	|\quad &x=\text{\ttfamily ClassAlloc}\langle\rho\rangle(y) \\
	|\quad &x=\text{\ttfamily GetContents}\langle\rho\rangle(y) \\
	|\quad &x=(\overline{y_i}) \\
	|\quad &x=y[i]
\end{align*}
we can express instruction normalization rules as in \cref{fig:instr-norm}.
\begin{figure*}
\begin{mathpar}
	\inferrule{y \rightsquigarrow y' \quad z = live_{c_i}(y')}{x=\text{\ttfamily ClassAlloc}\langle\tau_{c_i}\rangle(y)\rightsquigarrow x=(z, \text{\ttfamily Const}(i))}

	\inferrule{y \rightsquigarrow y'}{x=\text{\ttfamily ClassAlloc}\langle\tau\rangle(y)\rightsquigarrow x=\text{\ttfamily ClassAlloc}\langle\tau\rangle(y')}

	\inferrule{y \rightsquigarrow (y_0', y_1')}{x=\text{\ttfamily GetContents}\langle\tau\rangle(y)\rightsquigarrow x=y_0'}

	\inferrule{y \rightsquigarrow (y_0', y_1')}{x=\text{\ttfamily GetTag}\langle\tau\rangle(y)\rightsquigarrow x=y_1'}
\end{mathpar}
\caption{Normalization of SSA Instructions.}
\label{fig:instr-norm}
\end{figure*}

\subsection{Variant Equality}
Because ADT values have no identity, in Virgil they are compared with structural equality.
For unboxed variants, equality is normalized as a switch over the variant's tag, followed by comparing fields for each case.
Since ADT definitions may be (mutually) recursive, the generated equality routine can have (mutually) recursive calls.

\subsection{Packed Values}
For variant allocation, the compiler emits code that assembles packed scalars by performing bitwise operations based on the calculated intervals of each field.
Field access is similarly normalized as bitwise operations.

We have to be more careful when assembling scalars containing packed references.
In order to represent packed references, we introduce a new value type \lstinline|IntRepType| in the IR for bits backed by an integer, representing a specific scalar.
For instance, on a 64-bit target, a packed reference would be an \lstinline|IntRepType| backed by \lstinline|u64|, representing an \lstinline|R64| scalar.
Since the scalar is known, the machine lowering phase can tell if a value of \lstinline|IntRepType| represents a reference, in order to generate accurate bitmaps or stackmaps.

\subsection{Live Records}
Because Virgil's compilation model allows programs to run arbitrary code at compile time, a program's initialized heap may contain arbitrarily many variant values.
Because of desugaring to classes, these variant values are records that represent instances of these classes.
We modified the compiler's normalization pass to translate these records to their flattened representation as (multi-) scalars and substitute them where they occur.

\section{Conclusion and Future Work}

In this paper, we presented automated and partially-automated unboxing for Virgil ADTs.
Unlike previous work on unboxing for, e.g. OCaml, we leverage the whole program compilation approach of Virgil's compiler to address polymorphic types through specialization.
We introduce a notation for defining bit-precise packings of Virgil ADTs based on work from Titzer and Palsberg in 2005, which gives more expressivity than previous work.
Unboxing annotations and programmer declarations lead to packing problems that are not easily solved optimally by known algorithms.
In this paper we introduced a backtracking heuristic and scoring function that leads a reasonable heuristic.
At the time of this writing, unboxing is being stabilized in the Virgil compiler with intent to release to stable at the next available revision.
Early experiments show speedups for specific microbenchmarks from 1.2$\times$ to 5$\times$, though one benchmark shows a slowdown of 1.7$\times$.
Yet no microbenchmark shows an increase in memory consumption, and reductions can be more than 50\%.
Experiments on the Wizard engine, which contains a highly important ADT for the metavalue representation for its interpreter loop, showed up to 4\% improvement.
The Virgil compiler also uses ADTs internally, and applying unboxing to specific ADTs achieved a compiler bootstrap improvement of up to 1\% of execution time.

In future work, we plan to outline the details of our solving algorithm, explore better heuristics, and apply an existing ILP solver.
As always, more benchmarks and a more comprehensive evaluation are needed to better understand the improvements possible with these techniques.

\begin{acks}
  We graciously thank Jens for his many years of friendship and guidance.
  Jens was a fantastic advisor and the years 2001-2003 at Purdue and 2003-2007 at UCLA were magical.
  We thank him for putting together a great research lab with a great atmosphere and many fun times, overlapping with his other excellent students Ma Di, Tian Zhao, and Dennis Brylow (at Purdue), Krishna Nandivida, Vids Samantha, Christian Grothoff (at Purdue/UCLA), Fernando Magno Quintao Pereira and Jonathan Lee (at UCLA).
  Happy 60th birthday!
  We wish you many more years of health and happiness.
\end{acks}

\bibliographystyle{ACM-Reference-Format}
\bibliography{paper}

\end{document}